\documentclass{aa}
\usepackage{graphicx}
\begin{document}

\authorrunning{K\"apyl\"a et al.}
\titlerunning{Rotation dependence of the mixing length relations}

   \title{Local models of stellar convection II:}

   \subtitle{Rotation dependence of the mixing length relations}

   \author{P. J. K\"apyl\"a
	  \inst{1}$^{,}$\inst{2},
          M. J. Korpi
	  \inst{3},
	  M. Stix
	  \inst{2}
          \and
          I. Tuominen
	  \inst{1}$^{,}$\inst{4}
	  }

   \offprints{P. J. K\"apyl\"a\\
	  \email{petri.kapyla@oulu.fi}
	  }

   \institute{Astronomy Division, Department of Physical Sciences,
              PO BOX 3000, FIN-90014 University of Oulu, Finland
	  \and Kiepenheuer--Institut f\"ur Sonnenphysik, 
	      Sch\"oneckstrasse 6, D--79104 Freiburg, Germany
	  \and NORDITA, Blegdamsvej 17, DK-2100 Copenhagen, Denmark
	  \and Observatory, PO BOX 14, FIN-00014 University of Helsinki, 
	      Finland\\ }

   \date{Received  25 October 2004 / Accepted 29 March 2005}

   \abstract{We study the mixing length concept in comparison to
     three-dimensional numerical calculations of convection with
     rotation. In a limited range, the velocity and temperature
     fluctuations are linearly proportional to the superadiabaticity,
     as predicted by the mixing length concept and in accordance with
     published results. The effects of rotation are investigated by
     varying the Coriolis number, ${\rm Co} = 2\,\Omega \tau$, from
     zero to roughly ten, and by calculating models at different
     latitudes. We find that $\alpha$ decreases monotonically as a
     function of the Coriolis number. This can be explained by the
     decreased spatial scale of convection and the diminished
     efficiency of the convective energy transport, the latter of
     which leads to a large increase of the superadibaticity, $\delta
     = \nabla - \nabla_{\rm ad}$ as function of Co. Applying a
     decreased mixing length parameter in a solar model yields very
     small differences in comparison to the standard model within the
     convection zone. The main difference is the reduction of the
     overshooting depth, and thus the depth of the convection zone,
     when a non-local version of the mixing length concept is
     used. Reduction of $\alpha$ by a factor of roughly 2.5 is
     sufficient to reconcile the difference between the model and
     helioseismic results. The numerical results indicate reduction of
     $\alpha$ by this order of magnitude.

   }

   \maketitle


\section{Introduction}
   The mixing length concept was originally developed to deal with
   terrestrial incompressible convection (e.g. Taylor \cite{Taylor15};
   Prandtl \cite{Prandtl25}). Later, this approach was adopted in the
   context of stellar convection (Biermann \cite{Biermann1932};
   Cowling \cite{Cowling1935}), and the version introduced by Vitense
   (\cite{Vitense53}, see also B\"ohm-Vitense \cite{BohmVi58}) is
   still the most widely used description for convection in stellar
   structure models. The main advantage of this model is its
   simplicity; only one fixed free parameter is needed which relates
   the correlation length to the local pressure scale height. In
   stellar structure and evolution models the mixing length parameter,
   $\alpha$, is needed in order to determine the temperature gradient
   in the convection zone. The value of $\alpha$ fixes the depth of
   the convection zone in the model, which, for the solar case, is
   known from helioseismology (e.g. Monteiro et
   al. \cite{Monteiroea94}; Christensen-Dalsgaard et
   al. \cite{ChrisDalsea95}).

   Despite the crude nature of the mixing length concept, it can be
   useful in the bulk of the solar convection zone due to the fact
   that the temperature gradient there is nearly adiabatic, and the
   actual description of convection does not make much difference.

   One way to test the validity of the mixing length concept is to
   perform numerical calculations of convection. Comparisons have been
   done e.g. by Chan \& Sofia (\cite{ChanSo87}, \cite{ChanSo89}), Kim
   et al. (\cite{Kimea96}), and Porter \& Woodward
   (\cite{PoWo00}). The main result of these studies is that the
   relations derived between various mean thermodynamic and kinematic
   quantities under the basic mixing length assumption are good
   approximations when the superadiabaticity, $\delta = \nabla -
   \nabla_{\rm ad}$, is small enough (Kim et al. conclude that it is
   sufficient to have $\delta \leq 0.01$). Interestingly, Chan \&
   Sofia (\cite{ChanSo87}) find that the basic mixing length
   assumption, $l = \alpha H_{\rm p}$, is realised in their model.

\begin{figure}
   \centering
   \includegraphics[width=0.4\textwidth]{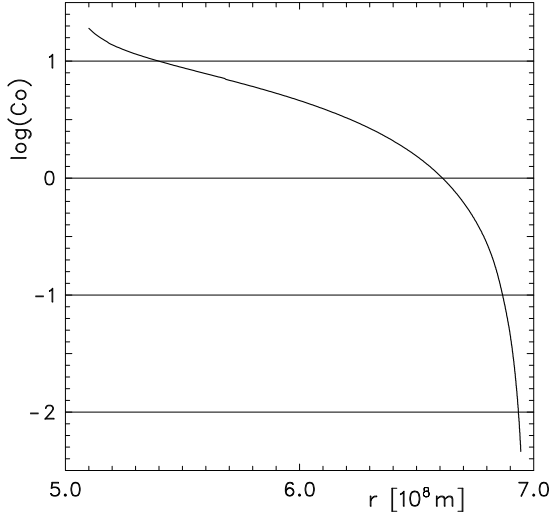}
      \caption{Logarithm of the Coriolis number, ${\rm Co} =
        2\,\Omega_\odot \alpha H_{\rm p}/v$, in the solar convection
        zone.}
      \label{fig:comlt}
   \end{figure}

   The mixing length concept, and the aforementioned comparisons to
   numerical convection, all neglect the effects of rotation, which
   can be dynamically important in many stars with convective
   envelopes. For example, the mixing length models of the solar
   convection zone (e.g. Stix \cite{Stix02}) yield velocities of the
   order of $10 {\rm m}\,{\rm s}^{-1}$ near the bottom of the
   convection zone. Furthermore, assuming the approximate size of the
   convective elements to be of the order of the pressure scale height
   ($H_{\rm p} \approx 5 \cdot 10^{7}{\rm m}$), one can estimate the
   turnover time to be $\tau \approx 10^{6}{\rm s}$. The solar angular
   velocity is $\Omega_{\odot} \approx 2.6 \cdot 10^{-6}{\rm s}^{-1}$,
   with which one can estimate the Coriolis number, ${\rm Co} =
   2\,\Omega \tau$, to be of order ten near the bottom of the
   convection zone and of order unity in a large fraction of the
   convection zone (Fig. \ref{fig:comlt}).

   We present three-dimensional calculations of convection in a
   rectangular domain that represents a small portion of a full star
   at some latitude. The latitude of the box can be chosen by imposing
   a suitably oriented rotation vector. In the present paper, our main
   aim is to parametrise the effects of rotation on the mixing length
   $\alpha$. This is achieved by calculating several boxes with
   varying rotational influence and latitude. One can think of the
   variation of the Coriolis number as the variation of the radial
   position in a convection zone. Thus, we obtain a radial profile for
   the mixing length $\alpha$ which we introduce into a solar model.

   In Sect.\,\ref{sec:model} of this paper the computational model is
   described, and in Sect.\,\ref{sec:mlt} a short summary of the
   mixing length concept is given. In Sects.\,\ref{sec:physnum} and
   \,\ref{sec:results} we discuss the parameter ranges of the
   calculations and summarise the results.


\section{The model}
\label{sec:model}
   A description of the model setup can be found in K\"apyl\"a et
   al. (\cite{Kapetal04a}; hereafter Paper I), and that of the
   numerical method in Caunt \& Korpi (\cite{CauKo2001}). The
   computational domain is a rectangular box, situated at a latitude
   $\Theta$.  The coordinates are chosen so that $x$, $y$, and $z$
   correspond to the south-north, west-east, and radially inward
   directions, respectively. The angular velocity as function of
   latitude is $\vec{\Omega} = \Omega (\cos \Theta \vec{\hat{{e}}_{x}}
   - \sin \Theta \vec{\hat{{e}}_{z}})$.

   The gas is assumed to obey the ideal gas equation
   \begin{eqnarray}
   p = \rho e (\gamma - 1)\;,
   \end{eqnarray}
    where $\gamma = c_{\rm P}/c_{\rm V} = 5/3$ is the ratio of the
   specific heats.

   To model the radiative losses near the surface we use a narrow
   cooling layer on the top of the convection zone, cooled with a term
   \begin{eqnarray}
   \Gamma_{\rm cool} = \frac{1}{t_{\rm cool}} f(z) (e - e_{0})\;,
   \label{equ:cool}
   \end{eqnarray}
   where $t_{\rm cool}$ is a cooling time, chosen to be short enough
   for the upper boundary to stay isothermal, $f(z)$ a function which
   vanishes everywhere else but in the interval $z_{0} \le z < z_{1}$,
   and $e_{0} = e(z_{0})$ the value of internal energy at the top of 
   the box.

   We adopt periodic boundary conditions in the horizontal directions,
   and closed stress free boundaries at the top and bottom. The
   temperature is kept fixed at the top of the box and a constant heat
   flux is applied at the bottom
   \begin{eqnarray}
   \frac{\partial u_{x}}{\partial z} = \frac{\partial u_{y}}{\partial z} 
   = u_{z} &=& 0\; \hspace{2cm} {\rm at} \;\;\;z = z_{0},z_{3}\;; \\
   e(z_{0}) &=& e_{0}\;, \\
   \frac{\partial e}{\partial z}\Big|_{z_{3}} &=& 
   \frac{g}{(\gamma - 1)(m_{3} + 1)}\;,
   \label{equ:dez}
   \end{eqnarray}
   where $m_{3}$ is the polytropic index in the lower overshoot layer.

   We obtain nondimensional quantities by setting
   \begin{eqnarray*}
   d = \rho_{0} = g = c_{\rm P} = 1\;.
   \end{eqnarray*}
   Thus, length is measured with respect to the depth of the unstable
   layer, $d = z_{2} - z_{1}$, and density in units of the initial
   value at the bottom of the convectively unstable layer,
   $\rho_{0}$. Time is measured in units of the free fall time,
   $\sqrt{d/g}$, velocity in units of $\sqrt{dg}$, and entropy in
   terms of $c_{\rm P}$. The box has horizontal dimensions $L_{x} =
   L_{y} = 4$, and $L_{z} = 2$ in the vertical direction. The upper
   boundary is situated at $z_0 = -0.15$ and the upper boundary of the
   convectively unstable layer is at $z_1 = 0$. The lower boundaries
   of the unstable layer and of the lower stably stratified region are
   at $(z_2,z_3)$ = (1,1.85), respectively.

   The dimensionless parameters controlling the calculations are the
   Prandtl number Pr, and the Taylor and Rayleigh numbers, denoted by
   Ta and Ra, respectively. The relative importance of thermal
   diffusion against the kinetic one is measured by the Prandtl number
   \begin{eqnarray}
     {\rm Pr} &=& \frac{\nu}{\chi_{0}}\;,
   \end{eqnarray}
   where $\chi_{0}$ is the reference value of the thermal diffusivity,
   taken from the middle of the unstably stratified layer. In the
   present study, $\nu$ is a constant. We set Pr = 0.4 which is a
   compromise between computational time (large diffusivity enforces a
   smaller time step) and realistic physics (in the Sun Pr $\ll 1$).

   Rotation is measured by the Taylor number
   \begin{eqnarray}
     {\rm Ta} = \Big( \frac{2\,\Omega d^{2}}{\nu} \Big)^{2}\;,
   \end{eqnarray}
   A related dimensionless quantity is the Coriolis number, which is
   the inverse of the Rossby number, Co = $2\,\Omega \tau$, where
   $\tau = d/u_{t}$ is the convective turnover time, and $u_{t}$ the
   rms-value of the fluctuating velocity determined through the
   calculation and averaged over the convectively unstable layer and
   time. In our calculations Co varies by about two orders of
   magnitude from about 0.1 to roughly 14.

   Convection efficiency is measured by the Rayleigh number
   \begin{eqnarray}
     {\rm Ra} = \frac{d^{4}g\delta}{\chi_{0}\nu H_{\rm p}}\;,
   \end{eqnarray}
   where $\delta = \nabla - \nabla_{\rm ad}$ is the superadiabaticity,
   measured as the difference between the actual and adiabatic
   logarithmic temperature gradients, and $H_{\rm p}$ the pressure
   scale height, both evaluated in the middle of the unstably
   stratified layer in the non-convecting hydrostatic reference
   solution.

   We define the Reynolds number as
   \begin{eqnarray}
     {\rm Re} = \frac{u_{t} d}{\nu}\;.
   \end{eqnarray}

   The initial stratification is polytropic, described by the indices
   $m_{1},\ m_{2}$, and $m_{3}$ for the three layers. We set $m_{1} =
   \infty,\ m_{2} = 1$, and $m_{3} = 3$, respectively, which means
   that the cooling layer is initially isothermal, and the
   stratification of the lower overshoot layer resembles that of the
   solar model of Stix (\cite{Stix02}).

   Initially the radiative flux, $\vec{F}_{\rm rad} = \kappa \nabla
   e$, where $\kappa = \gamma \rho \chi$ is the thermal conductivity,
   carries all of the energy through the domain. This constraint
   defines the thermal conductivities in each layer as
   \begin{eqnarray}
     \frac{\kappa_{i}}{\kappa_{j}} = \frac{m_{i} + 1}{m_{j} + 1}\;.
     \label{equ:kappa}
   \end{eqnarray}
   In the calculations the vertical profile of
   $\kappa$ is kept constant, which with the boundary condition for
   the internal energy, Eq.~(\ref{equ:dez}), assures that the heat
   flux into the domain is constant at all times. The initial state is
   perturbed with small scale velocity fluctuations of the order of
   $10^{-2} \sqrt{dg}$ which are deposited in the convectively
   unstable layer.

   The code is parallelised using a message passing interface
   (MPI). The calculations were carried out on the IBM eServer Cluster
   1600 supercomputer hosted by CSC Scientific Computing Ltd., in
   Espoo, Finland, and on the KABUL and BAGDAD Beowulf clusters with
   16 and 34 processors, respectively, at the Kiepenheuer-Institut
   f\"ur Sonnenphysik, Freiburg, Germany.

   \begin{table}
   \centering
      \caption[]{Summary of the convection calculations.}
      \vspace{-0.7cm}
      \label{tab:runs}
     $$
         \begin{array}{p{0.2\linewidth}cccccc}
            \hline
            \noalign{\smallskip}
            Run       & {\rm Ra} & {\rm Re} & {\rm Ta} & {\rm Co} & \Theta & {\rm Grid} \\
            \noalign{\smallskip}
            \hline
	    Co0       & 2.5\cdot 10^{5} & 140 & 0                 &    0 & -       & 64^3 \\
            \hline
	    Co01-00   & 2.5\cdot 10^{5} & 140 & 203               & 0.10 & \;\;0  \degr & 64^3\\
	    Co01-30   & 2.5\cdot 10^{5} & 140 & 203               & 0.10 & -30 \degr & 64^3\\
	    Co01-60   & 2.5\cdot 10^{5} & 138 & 203               & 0.10 & -60 \degr & 64^3\\
	    Co01-90   & 2.5\cdot 10^{5} & 138 & 203               & 0.10 & -90 \degr & 64^3\\
	    Co1-00    & 2.5\cdot 10^{5} & 139 & 2.03 \cdot 10^{4} & 1.04 & \;\;0  \degr & 64^3\\
	    Co1-30    & 2.5\cdot 10^{5} & 145 & 2.03 \cdot 10^{4} & 1.00 & -30 \degr & 64^3\\
	    Co1-60    & 2.5\cdot 10^{5} & 141 & 2.03 \cdot 10^{4} & 1.03 & -60 \degr & 64^3\\
	    Co1-90    & 2.5\cdot 10^{5} & 139 & 2.03 \cdot 10^{4} & 1.05 & -90 \degr & 64^3\\
	    Co10-00   & 2.5\cdot 10^{5} & 337 & 2.03 \cdot 10^{6} & 4.24 & \;\;0  \degr & 96^2 \times 64\\
	    Co10-30   & 2.5\cdot 10^{5} & 121 & 2.03 \cdot 10^{6} & 11.8 & -30 \degr & 96^2 \times 64\\
	    Co10-60   & 2.5\cdot 10^{5} & 105 & 2.03 \cdot 10^{6} & 13.6 & -60 \degr & 96^2 \times 64\\
	    Co10-90   & 2.5\cdot 10^{5} & 104 & 2.03 \cdot 10^{6} & 13.7 & -90 \degr & 96^2 \times 64\\
            \hline
	    s1Co0     & 3.1\cdot 10^{5} & 116 & 0             & 0 & - & 64^3\\
	    s2Co0     & 1.1\cdot 10^{6} & 250 & 0             & 0 & - & 128^3\\
	    s2Co01-90 & 1.1\cdot 10^{6} & 256 & 650               & 0.10 & -90 \degr & 128^3\\
	    s2Co1-00  & 1.1\cdot 10^{6} & 259 & 6.5 \cdot 10^4    & 1.00 &  0 \degr & 128^3\\
	    s2Co1-30  & 1.1\cdot 10^{6} & 265 & 6.5 \cdot 10^4    & 0.98 & -30 \degr & 128^3\\
	    s2Co1-60  & 1.1\cdot 10^{6} & 262 & 6.5 \cdot 10^4    & 1.00 & -60 \degr & 128^3\\
	    s2Co1-90  & 1.1\cdot 10^{6} & 263 & 6.5 \cdot 10^4    & 1.00 & -90 \degr & 128^3\\
	    s2Co10-90 & 1.1\cdot 10^{6} & 221 & 6.5 \cdot 10^6    & 12.4 & -90 \degr & 128^3\\
	    s3Co0 & 1.8\cdot 10^{6} & 340 & 0             & 0 & - & 192^3\\
	    s4Co0 & 4.1\cdot 10^{6} & 609 & 0             & 0 & - & 256^3\\
            \noalign{\smallskip}
            \hline
         \end{array}
     $$
   \end{table}

   \begin{figure*}
   \centering
   \includegraphics[width=0.96\textwidth]{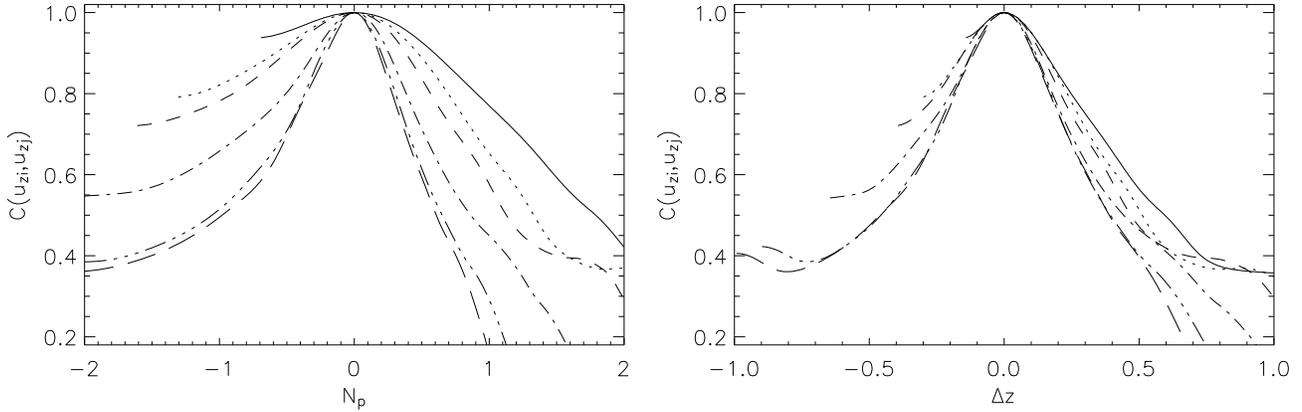}
      \vspace{-6.5cm}
      \caption{Correlations of $u_z$ with respect to the reference
        levels $z_{\rm ref} = 0$ (solid), 0.17 (dotted), 0.26
        (dashed), 0.5 (dot-dashed), 0.77 (triple dot dashed), and 0.87
        (long dashed) from the run s2Co0 as function of the number of
        pressure scale heights, $N_{\rm p}$, (left) and the actual
        distance $\Delta z$ (right).}
      \label{fig:corle}
   \end{figure*}


\section{The mixing length concept}
\label{sec:mlt}
   The basic assumption in the mixing length concept is that the
   rising and descending convective elements preserve their identity
   for a length which is proportional to the local pressure scale
   height, i.e.
   \begin{eqnarray}
     l = \alpha H_{\rm p}\;
     \label{equ:basicamlt}
   \end{eqnarray}
   where $H_{\rm p} = \big( \frac{1}{p}\frac{\partial p}{\partial z}
   \big)^{-1}$. Using this assumption it is possible to derive
   equations for the mean convective velocity and temperature
   fluctuation (e.g. Stix \cite{Stix02})
   \begin{eqnarray}
     \langle u_{z}'^{2} \rangle & = & \frac{\alpha_{\rm u}^{2}H_{\rm p}g}{8} 
       (\nabla - \nabla_{\rm ad}) \;,  \label{equ:vzmlt} \\
     \langle \sqrt{T'^{2}} \rangle & = & \frac{\alpha_{\rm T}}{2} 
       (\nabla - \nabla_{\rm ad}) \langle T \rangle \;,
   \end{eqnarray}
   where adiabatic variation of the convective eddies has been assumed,
   and the brackets denote horizontal and temporal averaging.

   The mixing length parameter enters the stellar evolution models in
   the context of energy transport. Thus, we relate the fluxes of
   enthalpy and kinetic energy with the additional equations
   (e.g. Porter \& Woodward \cite{PoWo00})
   \begin{eqnarray}
     F_{\rm e} & = & \alpha_{\rm e} c_{\rm p} \langle \rho \rangle \langle \sqrt{T'^{2}} \rangle \langle \sqrt{u_{z}'^{2}} \rangle\;, \\
     F_{\rm kin} & = & \alpha_{\rm k} \langle \rho \rangle \langle \sqrt{u_{z}'^{2}} \rangle^{3/2}\;, \label{equ:fkinmlt} 
   \end{eqnarray}
   where $F_{\rm e} = \langle c_{\rm p} \rho T' u_{z}' \rangle$ and
   $F_{\rm kin} = \langle \frac{1}{2} \rho \vec{u}^{2} u_{z}'\rangle$
   are the enthalpy and kinetic energy fluxes, respectively. We use
   subscripts u, T, e, and k for the mixing length parameters obtained
   from the relations (\ref{equ:vzmlt}) to (\ref{equ:fkinmlt}) due to
   the fact that it is not \emph{a priori} clear that all of them
   would yield the same value.

   \begin{figure}
   \centering
   \includegraphics[width=0.425\textwidth]{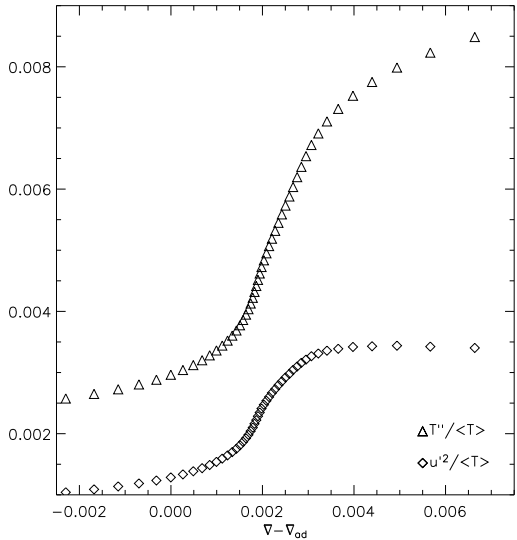}
      \caption{Relations between the vertical velocity and temperature
        fluctuations versus the superadiabatic gradient from the run
        s2Co1-90.}
      \label{fig:calpha}
   \end{figure}

\section{Parameter ranges in the Sun and in the calculations}
\label{sec:physnum}

  \subsection{Coriolis number in the Sun}
   The Coriolis number varies strongly in the outer convection zone of
   the Sun. Near the surface it is only of order $10^{-3}$ or smaller,
   but it exceeds 10 in the deeper region. Figure~\ref{fig:comlt}
   shows Co for a standard solar model, calculated according to ${\rm
   Co} = 2\,\Omega l/v$, where $l = \alpha H_{\rm p}$ is the mixing length,
   and $v$ is the convection velocity that is obtained by means of the
   local mixing-length formalism. The
   coefficient $\alpha =1.66$ is obtained through a calibration of a
   model of age $4.57\cdot 10^9$~years to the present Sun.

   \subsection{Summary of the calculations}
   Table \ref{tab:runs} gives a summary of the calculations. Bearing
   in mind the result presented in Fig. \ref{fig:comlt}, we have made
   calculations with four different Taylor numbers, which roughly
   correspond to Coriolis numbers 0, 0.1, 1, and 10. We refer to these
   runs with a prefix Co0, Co01, Co1, and Co10, respectively. If we
   interprete the Coriolis number as the depth dependence, the two
   first cases represent roughly the top, and the latter two the
   middle and the bottom of the solar convection zone. To investigate
   the latitudinal dependence we have made calculations at four
   latitudes, namely $\Theta = 0 \degr$ (equator), $-30 \degr$, $-60
   \degr$, and $-90 \degr$ (southern pole). In the standard setup the
   essentially step-function form of the thermal conductivity profile,
   see Eq.~(\ref{equ:kappa}), introduces a peak in the
   superadiabaticity at the bottom of the convection zone (not
   shown). We find this feature to persist in calculations where the
   resolution is doubled from the values listed in Table
   \ref{tab:runs} using runs from Paper I.

   Thus we introduce a setup in which the logarithmic temperature
   gradient in the lower overshoot layer in the initial state follows
   a profile
   \begin{eqnarray}
     \nabla = \nabla_{\rm 3} + \frac{1}{2} \{\tanh [4(z_{\rm m} - z)] +
     1\}\Delta \nabla\;,
   \end{eqnarray}
   where $\nabla_3$ is the applied gradient at the bottom, and $\Delta
   \nabla$ the difference between the gradient in the unstable layer
   and the applied gradient, and $z_{\rm m}$ inflection point of the
   tanh-function, calculated so that $\nabla = \nabla_{\rm ad}$ at the
   interface between the stable and unstable layers. The thermal
   conductivity is determined so that initially radiative diffusion
   transports the total energy flux. We denote runs with this setup by
   a prefix s and a number which depends on the used resolution in
   Table \ref{tab:runs}.

   \begin{figure*}
   \centering
   \includegraphics[width=1.\textwidth]{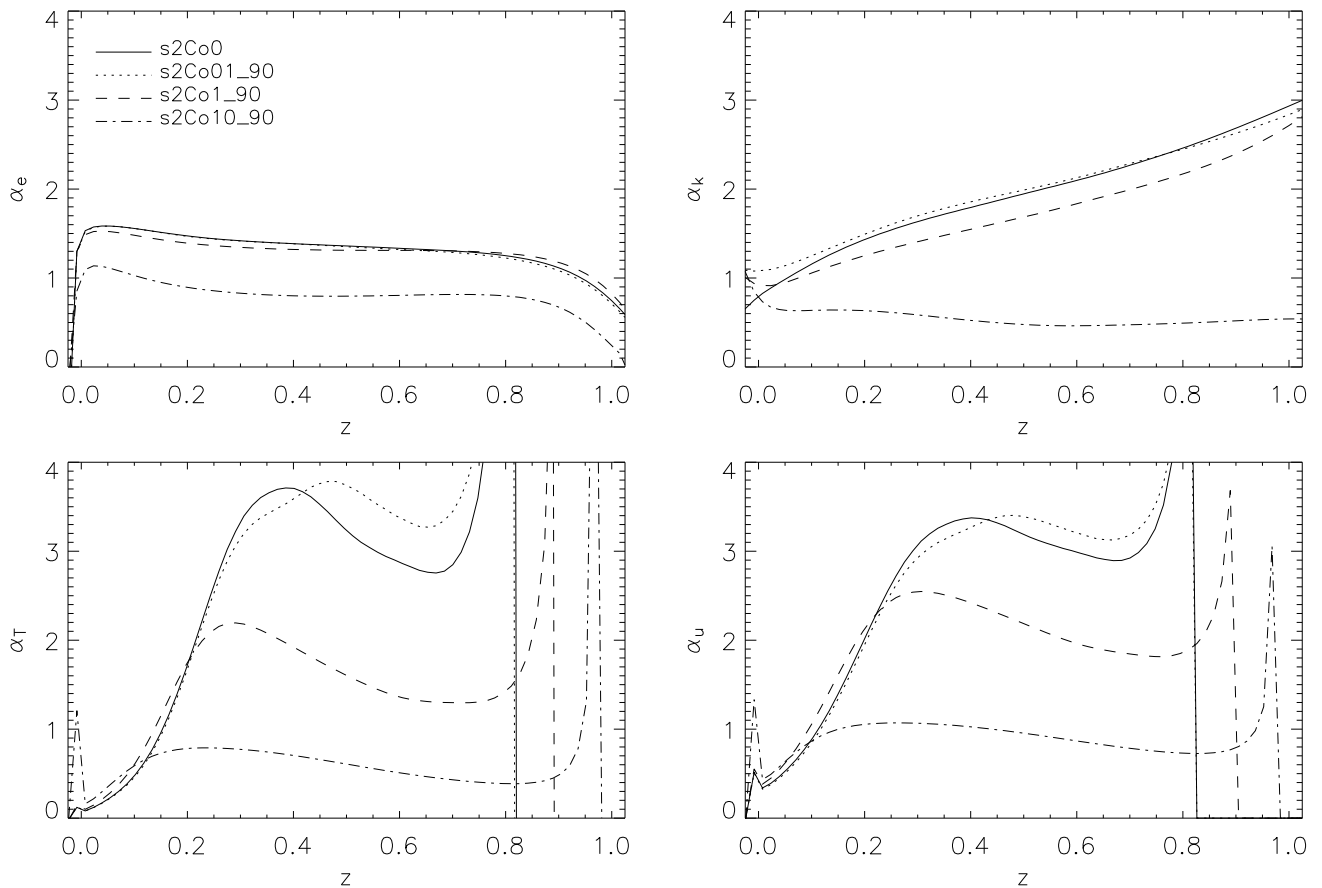}
      \caption{The mixing length parameters calculated from
        Eqs.~(\ref{equ:vzmlt}) to (\ref{equ:fkinmlt}) as functions of
        depth and rotation.}
      \label{fig:alpha}
   \end{figure*}
 

\section{Results}
\label{sec:results}

\subsection{Basic mixing length assumption}
   We test the basic mixing length assumption by calculating the
   correlation of the horizontally averaged vertical velocity as
   function of depth as suggested by Chan \& Sofia (\cite{ChanSo87},
   \cite{ChanSo89})
   \begin{eqnarray}
     C[u_{z_{\rm ref}},u_{z}] = \frac{\langle u_{z_{\rm ref}} u_{z}
       \rangle}{\langle u_{z_{\rm ref}}^2 \rangle^{1/2} \langle
       u_{z}^2 \rangle^{1/2}}\;,
   \end{eqnarray}
   where $z_{\rm ref}$ is the reference level and $z$ runs from $z_0$
   to $z_3$. It is useful to plot this quantity as a function of the
   number of pressure scale heights
   \begin{eqnarray}
     N_{\rm p} = \Delta \ln \langle p \rangle\;,
   \end{eqnarray}
   where the difference is calculated from the reference level. If the
   basic mixing length assumption, defined by
   Eq.~(\ref{equ:basicamlt}), is correct the correlations plotted as
   functions of $N_{\rm p}$ with respect to different reference levels
   should fall on top of each other. Figure \ref{fig:corle} shows the
   correlations in the run s2Co0 case as functions of $N_{\rm p}$ and
   the distance measured in terms of $d$. The widths of the
   correlations measured as function of pressure scale heights
   decrease monotonically as a function of depth. We find that this
   width is more or less constant when measured in terms of $d$, the
   unit length. This result resembles the one by Robinson et
   al. (\cite{Robinsonea03}), who present calculations of convection
   representing the highly superadiabatic near-surface layers of the
   Sun.

   Although we do not find a scaling of the correlation width with the
   scale height reported by Chan \& Sofia (\cite{ChanSo87},
   \cite{ChanSo89}), we do find that the velocity and temperature
   fluctuations follow a more or less linear relation as function of
   the superadiabaticity, $\delta$, for a limited range of parameters
   (see Figs. \ref{fig:calpha} and \ref{fig:alpha}). This is realised
   in in the calculations where $\delta$ decreases monotonically
   within the convectively unstable region due to the more smoothly
   varying thermal conductivity (Figs. \ref{fig:supera} and
   \ref{fig:superalat}).

   The linear part in Figs. (\ref{fig:calpha}) and (\ref{fig:alpha})
   is clearly shorter than in the results of Chan \& Sofia
   (\cite{ChanSo87}, \cite{ChanSo89}, \cite{ChanSo96}) which can be
   explained by the fact that whereas in our calculations the
   convectively unstable region contains less than two pressure
   scale-heights, Chan \& Sofia typically have five or more. Our
   failure to find any such linear relation for the standard models is
   most probably due to an unrealistic step function like thermal
   conductivity profile which somewhat distorts the thermal
   structure. However, the qualitative trends seen as function of
   rotation do not depend on the choice of the thermal conductivity
   profile.

\subsection{Effect of rotation and dependence on latitude}
   Figure \ref{fig:alpha} shows the mixing length parameters, as
   defined by Eqs.~(\ref{equ:vzmlt}) to (\ref{equ:fkinmlt}), from the
   s2 runs with four Coriolis numbers at latitude $-90\degr$. These
   figures show that we do not find one universal parameter $\alpha$
   that could describe all the quantities in Eqs.~(\ref{equ:vzmlt}) to
   (\ref{equ:fkinmlt}). Firstly, $\alpha_{\rm e}$ describing the
   convective energy transport, seems to have a rather well-defined
   value of about 1.5 for all cases bar the runs with largest
   rotation.

\begin{figure}
   \centering
   \includegraphics[width=0.5\textwidth]{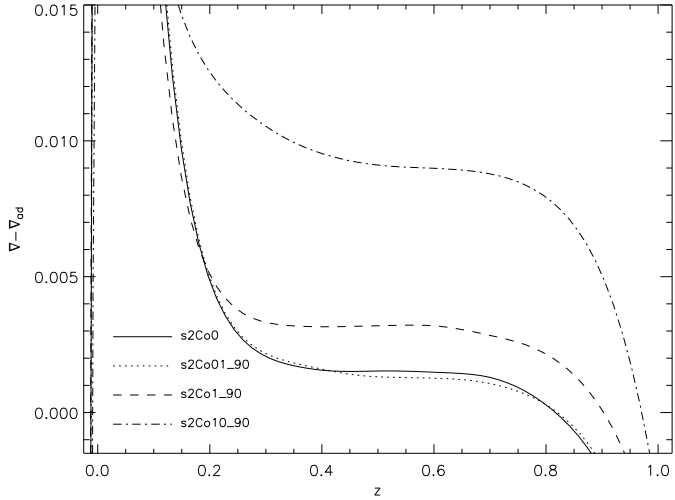}
      \caption{Superadiabaticity $\nabla - \nabla_{\rm ad}$ as
        function of depth and rotation.}
      \label{fig:supera}
   \end{figure}

   Similarly, $\alpha_{\rm k}$ stays fairly constant for slow
   rotation, and shows a large reduction for the Co10 case. On the
   other hand, $\alpha_{\rm T}$ and $\alpha_{\rm u}$ decrease rapidly
   as Co grows. The explanation to this behaviour is that the
   superadiabaticity, which appears in the denominator in the
   equations for $\alpha_{\rm T}$ and $\alpha_{\rm u}$, increases
   rapidly as a function of rotation (see Fig.
   \ref{fig:supera}). This growth of $\nabla - \nabla_{\rm ad}$ is
   linked to the decrease of the convective energy transport as a
   function of rotation (see Paper I). The reduced convective flux
   forces the radiative diffusion to transport more energy, which can
   only be achieved by steepening the temperature gradient. However,
   due to the fact that the present numerical model deals with
   inefficient convection, where maximally about one third of the
   total energy flux is transported by convection, the rotational
   influence on the superadiabaticity is probably underestimated in
   comparison to the solar case where convection is expected to
   transport practically all energy. Preliminary numerical results
   with an efficient convection setup support this conjecture.

\begin{figure}
   \centering
   \includegraphics[width=0.5\textwidth]{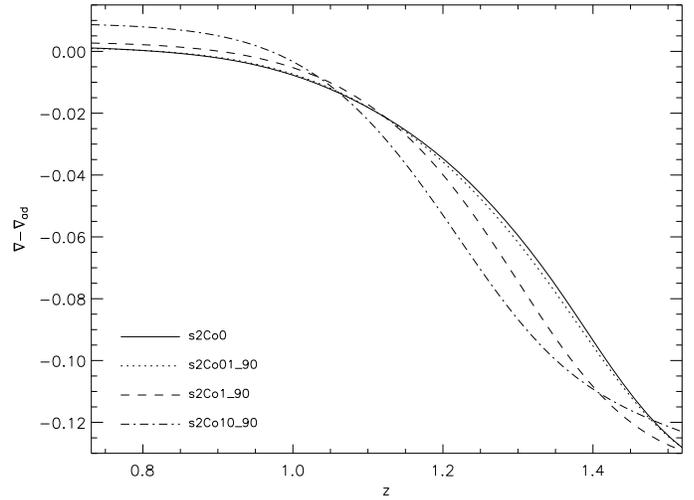}
      \caption{Superadiabaticity $\nabla - \nabla_{\rm ad}$ as a
        function of depth and rotation in the transition layer between
        the convectively unstable and stable regions. Note that the
        scale is different by roughly an order of magnitude in
        comparison to Fig. \ref{fig:supera}.}
      \label{fig:supera_over}
   \end{figure}

   Fig. \ref{fig:supera_over} shows the superadiabaticity in the
   overshoot layer from the s2 set of calculations. The increasing
   effect of rotation is seen in the larger magnitude of the
   superadiabaticity and a significantly deeper superadiabatic
   layer. Furthermore, as the overshooting decreases as a function of
   rotation the transition at the unstable/stable interface becomes
   steeper due to the larger $\delta$ in the convectively unstable
   layer. The quantitative change of $\delta$ in the depth of the
   superadiabatic layer is probably exaggerated due to the much larger
   energy flux in the present calculations in comparison to the Sun. A
   similar issue concerns the transition from the overshoot to the
   radiative layer noted by Rempel (\cite{Rempel2004}). According to
   that study, reducing the input flux sufficiently one should reach a
   regime where the overshoot resembles that obtained by non-local
   mixing length models, i.e. almost adiabatic overshoot and a sharp
   transition to the radiative gradient. However, this may require the
   use of an anelastic code due to the timestep requirement.

   \begin{figure}
   \centering
   \includegraphics[width=0.5\textwidth]{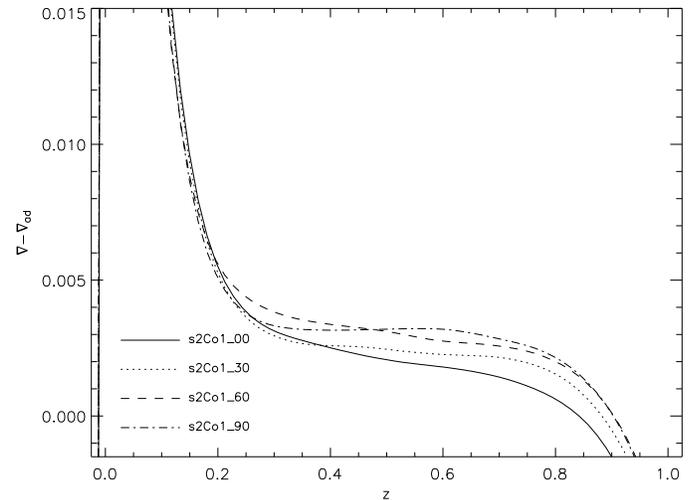}
      \caption{Superadiabaticity $\nabla - \nabla_{\rm ad}$ as
        function of depth and latitude.}
      \label{fig:superalat}
   \end{figure}

   \begin{figure*}
   \centering
   \includegraphics[width=1.\textwidth]{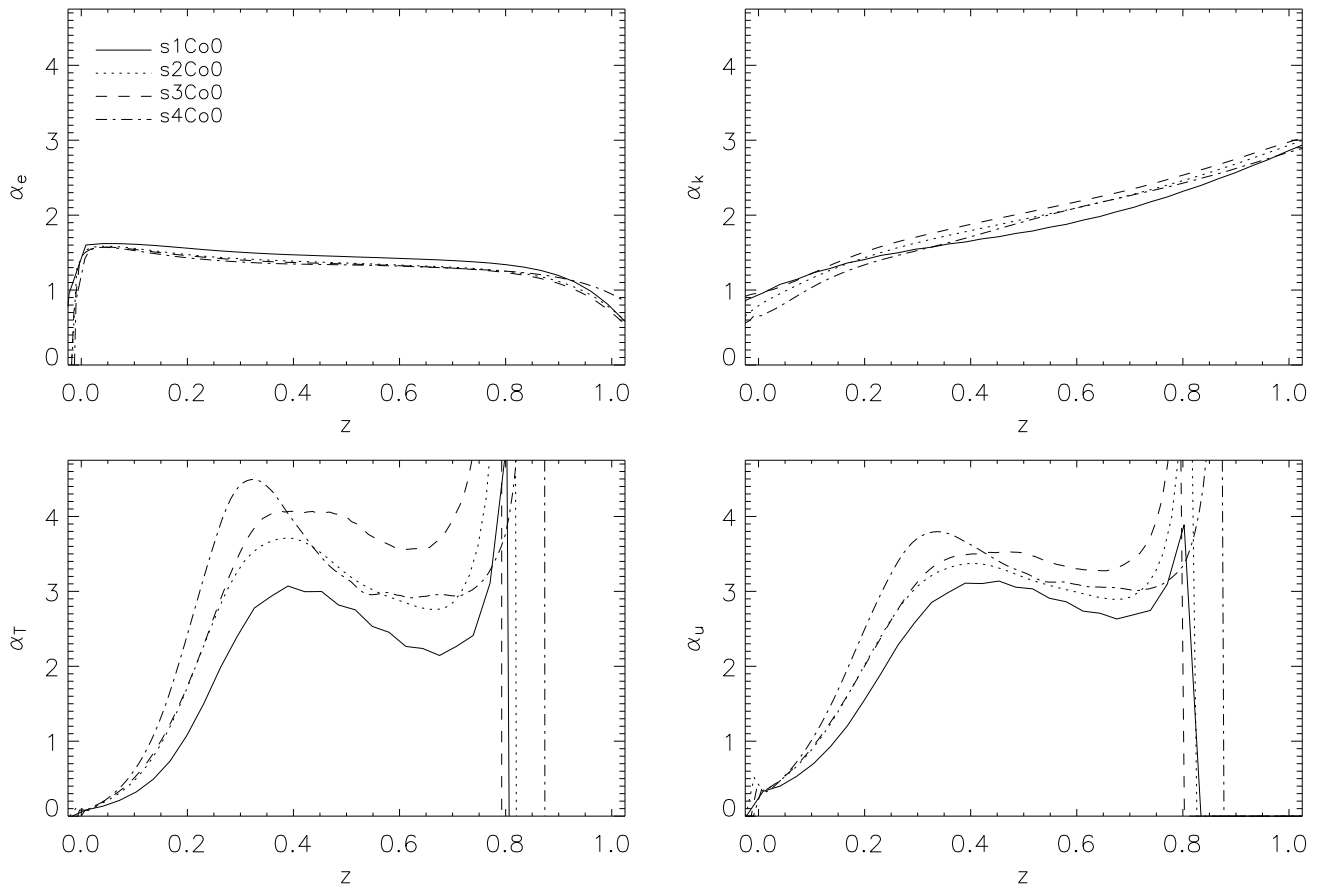}
      \caption{Mixing length parameters $\alpha_{\rm i}$, defined via
        Eqs.~(\ref{equ:vzmlt}) to (\ref{equ:fkinmlt}), as function of
        Ra and resolution for the runs siCo0 from Table
        \ref{tab:runs}.}
      \label{fig:alpha_Ra}
   \end{figure*}

   The main feature of the latitudinal dependence is that the
   superadiabaticity increases as function of latitude (see Fig.
   \ref{fig:superalat}). Note that whereas the superadiabatic region
   extends as the latitude increases, the overshooting depth decreases
   in a similar fashion to Fig. \ref{fig:supera_over}.  In an earlier
   study with denser latitude coverage (Paper I) some indications of a
   maximum of $\delta$ at latitudes $\Theta \approx -60\degr \ldots
   -75\degr$ was observed. What this implies is that convection is
   more efficient at low latitudes and that due to this one would
   expect overshooting to peak there. This is indeed observed for slow
   and moderate rotation, i.e. ${\rm Co} < 4$, see Fig. 9 of Paper
   I. This could mean that the overshooting extends deeper at
   equatorial regions, which can lead to a prolate shape of the
   tachocline as indicated by helioseismology (Basu \& Antia
   \cite{BasuAntia2001}). However, one must bear in mind that the
   helioseismology results refer to the shear layer whose connection
   with overshooting is not yet known.  The thermal stratification of
   the Sun does not vary observably as a function of latitude as in
   the numerical models which is most likely due to the much too
   vigorous convection in the calculation. Furthermore, we note that
   the trend in the overshooting depth as function of latitude seems
   to reverse for more rapid rotation (Paper I).

   We feel that further study, beyond of the scope the present paper,
   of the details of the overshooting as functions of rotation and
   input energy flux is needed in order to determine whether it is
   possible to approach the mixing length regime with 3D calculations
   and to fully substantiate the effects of rotation which are
   virtually always neglected in overshooting models.

   \subsection{Effects of resolution}
   In order to study the dependence of the mixing length relations on
   resolution, we have made a set of runs where we vary the Rayleigh
   number from $3.1 \cdot 10^5$ to $4.1 \cdot 10^6$ and the number of
   gridpoints from 64$^3$ to 256$^3$, named s1Co0 to s4Co0 in Table
   \ref{tab:runs}. Fig. \ref{fig:alpha_Ra} shows the mixing length
   parameters from the runs s1Co0 to s4Co0. The main feature of the
   results is that the qualitative character of the
   $\alpha$-parameters remains unaffected when resolution changes. The
   relatively large changes in the parameters $\alpha_{\rm T}$ and
   $\alpha_{\rm u}$ are mostly explained by the smaller
   superadiabaticity as resolution increases.

   The largest resolution calculation, however, is not fully
   comparable to the lower resolution ones due to the fact that the
   calculation did not reach a thermally saturated state. This has to
   do with the fact that in order to avoid the long thermal relaxation
   phase we start the calculation with a stratification that is
   expected to be close to the final thermally relaxed state. This
   procedure has to be carried out by trial and error for each
   resolution and for the 256$^3$ run the initial guess of the
   stratification was too superadiabatic, resulting in too efficient
   convection. However, we think that even though this run is not
   quantitatively fully comparable to the others, it still illustrates
   the fact that the qualitative results remain unchanged as a
   function of resolution.

   \begin{figure}[h]
     \centerline{\includegraphics[width=80mm]{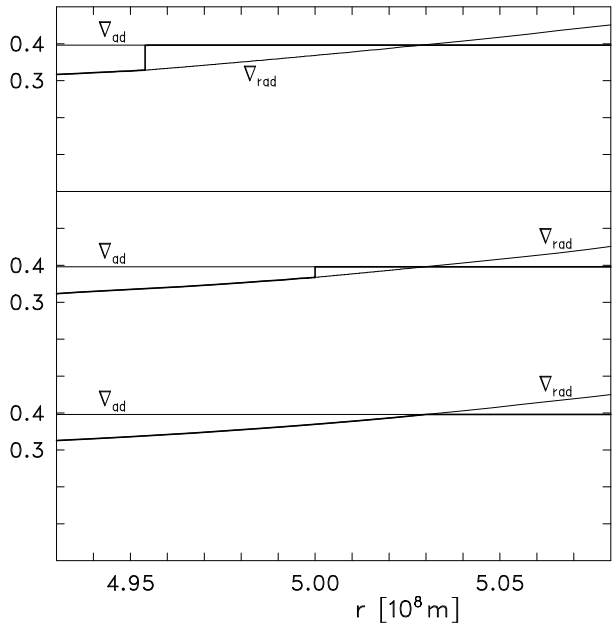}}
     \caption{Temperature gradients $\nabla = d\ln T/d \ln P$ near the
       base of the convection zone.  The actual gradient ({\em heavy
         line}) is identical to the radiative gradient in the core,
       and closely follows the adiabatic in the convection zone and in
       the region of overshooting. A local model with no overshooting
       ({\em lower line}); non-local models with mean path $l/2$ ({\em
         above}), as assumed in local theory, and with mean path $l/5$
       ({\em middle}).}
     \label{over}
   \end{figure}

\subsection{Application to a solar model}
   As the Coriolis number in the lower part of the convection zone is
   large, the convective efficiency should be substantially reduced
   according to the 3D numerical calculations. In the one-dimensional
   standard solar model we simulate this by a reduced mixing length.
   However, there are virtually no consequences to the solar structure
   {\em within} the convection zone. This is because the temperature
   gradient, as determined by mixing-length theory, closely follows
   the adiabatic gradient, irrespective of the details of the
   convection formalism.

   On the other hand, the overshooting {\em below} the base of the
   convection zone may be severely reduced by a reduced convective
   efficiency. This can be illustrated by solar models that employ a
   non-local version of the mixing-length formalism (Skaley \& Stix
   \cite{SkaStix91}, Stix \cite{Stix02}). In those models the
   temperature excess $\Delta T$ and the convection velocity $v$ are
   determined as integrals over the distance a convection parcel
   travels in the vertical direction. If this distance is reduced, the
   parcel acquires less kinetic energy, so that there is less
   overshooting. In Fig.~\ref{over} this is shown for a standard model
   calculated with local mixing-length theory and two non-local
   models. We find that the depth of overshooting decreases
   approximately in proportion to the decrease of the integration
   path. If the average parcel path is reduced by a factor 2.5, which
   is consistent with the numerical results, overshooting extends only
   over $\approx 2900$\,km.

   Christensen-Dalsgaard et al. (\cite{ChrisDalsea95}) determined an
   upper limit of $\approx 0.1H_{\rm p}$ for overshooting with a sharp
   transition from the adiabatic to the radiative regime, such as we
   obtain with the non-local mixing length models. Beyond this limit
   the discontinuity of the temperature gradient would leave a
   measurable signature in the spacing of p-mode frequencies. Thus,
   the rotational quenching of the overshooting would appear to be
   welcome. On the other hand, the total depth of the convection zone
   may become too small in this way. The problem is aggravated as a
   recent redetermination of the abundance of oxygen and other heavy
   elements (Asplund et al. \cite{Asplundea2004}) lowers the opacity
   and, therefore, also leads to a shallower convection zone. Bahcall
   et al. (\cite{Bachallea2005}) find $r/r_\odot = 0.726$ for the base
   of the convection zone of their model BP04+ which incorporates the
   new lower opacity as well as best available updates of all other
   input parameters. This means that the convection zone would be too
   shallow by $\approx 9000$\,km or $\approx 0.2H_{\rm p}$, as
   compared to the helioseismological result ($r/r_\odot = 0.713\pm
   0.001$; Christensen-Dalsgaard et al. \cite{ChrDalea1991}, Basu \&
   Antia \cite{BasuAntia1997}).

   Presently is is not clear how this conflict will be resolved.  New
   opacity calculations (Seaton \& Badnell \cite{SeatonBadnell2004})
   yield an increase of 5\,\% compared to the latest OPAL values; but
   this is only about half the increase needed: Bahcall et
   al. (\cite{Bachallea2005}) obtain a sufficiently deep convection
   zone if they increase the opacity by $\approx 10$\,\% in the depth
   range $0.4r_\odot$ to $0.7r_\odot$. We suggest that the solution is
   a combination of modest overshooting and a modest opacity
   increase. Overshooting may also account for the whole difference if
   it avoids the sharp transition that occurs in our
   calculation. Rempel (\cite{Rempel2004}) has proposed a model in
   which the mixing between downflows and upflows is crucial for the
   total depth of the overshooting, and in which an ensemble of
   downflows with different strength produces a smooth transition.

  
\begin{acknowledgements}
PJK acknowledges the financial support from the Finnish graduate
school for astronomy and space physics and the Kiepenheuer-Institut
for travel support. MJK acknowledges the hospitality of LAOMP,
Toulouse and the Kiepenheuer-Institut, Freiburg during her visits, and
the Academy of Finland project 203366. Travel support from the Academy
of Finland grant 43039 is acknowledged. The authors thank Mathieu
Ossendrijver for many helpful discussions and the anonymous referee
whose comments and suggestions greatly improved the manuscript.
\end{acknowledgements}

\end{document}